\pdfoutput=1
\documentclass[conference]{IEEEtran}
\IEEEoverridecommandlockouts

\usepackage{amsmath,amssymb,amsthm}
\usepackage{mathrsfs}
\usepackage{latexsym}
\usepackage{bm}
\usepackage{cases}

\usepackage{graphicx}

\usepackage{color}
\usepackage{accents}
\usepackage{enumerate}
\usepackage{cite}
\usepackage{stfloats}

\makeatletter
\def\widebar{\accentset{{\cc@style\underline{\mskip10mu}}}}
\def\Widebar{\accentset{{\cc@style\underline{\mskip8mu}}}}
\makeatother

\allowdisplaybreaks
\theoremstyle{plain}

\theoremstyle{definition}
\theoremstyle{definition} 

\usepackage[
  paper=letterpaper,
  top=0.7in,       
  bottom=1.0in,    
  left=0.62in,     
  right=0.62in,
  columnsep=0.24in 
]{geometry}

\def\BibTeX{{\rm B\kern-.05em{\sc i\kern-.025em b}\kern-.08em
    T\kern-.1667em\lower.7ex\hbox{E}\kern-.125emX}}
\begin{document}
\title{Differential Spatial Modulation with Transmit Diversity for Pinching-Antenna Systems}
\author{\IEEEauthorblockN{Yiwei~Tao\IEEEauthorrefmark{1}\IEEEauthorrefmark{4}, Yao~Ge\IEEEauthorrefmark{2}, Dong~Li\IEEEauthorrefmark{3}, Miaowen~Wen\IEEEauthorrefmark{1}, and Merouane~Debbah\IEEEauthorrefmark{4}
}
	\IEEEauthorblockA{\IEEEauthorrefmark{1}School of Electronic and Information Engineering, South China University of Technology, Guangzhou 510641, China}
	\IEEEauthorblockA{\IEEEauthorrefmark{2}AUMOVIO-NTU Corporate Lab, Nanyang Technological University, 639798, Singapore}
	\IEEEauthorblockA{\IEEEauthorrefmark{3}School of Computer Science and Engineering, Macau University of Science and Technology, Macau, China}
	\IEEEauthorblockA{\IEEEauthorrefmark{4}Research Institute for Digital Future, Khalifa University, 127788 Abu Dhabi, UAE}
		Email: eeyiweitao@mail.scut.edu.cn, yao.ge@ntu.edu.sg, dli@must.edu.mo, eemwwen@scut.edu.cn, merouane.debbah@ku.ac.ae 
\vspace{-0.2cm}}


\maketitle

\thispagestyle{empty}
\pagestyle{empty}

\vspace{-1.8cm}
\begin{abstract}
Pinching antenna (PA) systems provide a new spatial degree of freedom by flexible activation of pinching positions. 
However, the resulting effective channel strongly depends on the activated pinching positions, rendering conventional coherent transmission generally relies on accurate acquisition of instantaneous channel state information (CSI) and incurring substantial pilot overhead.
To address this challenge, we propose a differential spatial modulation (DSM) scheme for PA systems, termed as {\em DSM-PA}.
Specifically, a differential transmission scheme with an embedded Alamouti coding structure is designed, where information bits are conveyed via phase variations between adjacent symbol blocks. This design enables noncoherent transmission without requiring instantaneous CSI while simultaneously achieving transmit diversity. Moreover, to fully exploit the spatial degrees of freedom of PA systems, a pinching position-based index modulation (IM) rule is developed to enhance spectral efficiency.
An asymptotically tight upper bound on the average bit error rate (BER) over quasi-static Rician fading channels is derived using the moment-generating function (MGF) method. The diversity analysis also reveals that the proposed DSM-PA scheme achieves full transmit diversity.
Finally, simulation results verify the accuracy of the BER analysis and demonstrate the effectiveness of the proposed DSM-PA scheme.
\end{abstract}

\section{Introduction}
With the growing demand for flexible coverage, reduced hardware complexity, and highly reliable transmission in next-generation wireless communication systems, reconfigurable wireless transmission architectures have received considerable attention~\cite{11169486}.
Unlike conventional fixed antenna (FA) arrays, these architectures aim to exploit additional spatial degrees of freedom to reconfigure the wireless propagation environment, thereby improving the reliability and efficiency of wireless links.
Driven by these efforts, a variety of emerging paradigms have been developed, including reconfigurable intelligent surfaces (RIS), fluid antenna systems (FAS), and movable antenna systems (MAS)~\cite{8741198,9264694,10286328}. Among them, the pinching antenna (PA) system, as a novel waveguide-based flexible antenna architecture, has recently emerged as a promising paradigm and attracted increasing attention~\cite{10945421}.

In PA systems, the radio-frequency (RF) signal propagates along a dielectric waveguide with low loss. It is subsequently radiated into free space via small dielectric elements, i.e., pinching antennas, attached at specific locations along the waveguide. By flexibly deploying or activating pinching antennas at different positions, PA systems can form a reconfigurable distributed radiating aperture, thereby introducing a new degree of freedom in the pinching position domain for wireless transmission design. Besides, the scale of PA systems can be flexibly adjusted by adding or releasing pinching antennas along one or multiple waveguides, enabling a favorable tradeoff between system flexibility and hardware cost~\cite{11212813}.
Leveraging these features, PA systems provide a natural platform for pinching-position domain signal design and facilitate integration with index modulation (IM) and spatial modulation (SM) schemes to improve spectral efficiency and transmission~reliability~\cite{10845819}.

Nevertheless, the advantages of PA systems come at the cost of new challenges. The effective channel is strongly coupled with the activated pinching positions, such that different position patterns correspond to distinct channel responses. Consequently, coherent transmission relies on accurate instantaneous channel state information (CSI) at the receiver. When the pinching positions vary across transmission blocks, continuous channel estimation and tracking are required, resulting in substantial pilot overhead~\cite{11018390}.
Unfortunately, existing studies on PA systems have mainly focused on coherent transmission designs based on instantaneous CSI. For example, prior works have investigated the performance gains of PA systems from various perspectives, including pinching position optimization, integrated sensing and communications (ISAC), multiuser access~\cite{11122551,11478635,10912473}. In addition, IM has been incorporated into PA systems, where different pinching-position patterns are exploited to convey additional information and improve spectral efficiency~\cite{11368709}.
Although these methods achieve desired performance under ideal CSI conditions, their effectiveness heavily depends on the accuracy of channel estimation.
On the other hand, differential modulation and noncoherent SM have been extensively studied in conventional multi-antenna systems, and have been shown to enable reliable transmission without requiring instantaneous CSI~\cite{6879496}. However, existing differential and noncoherent schemes are primarily designed for conventional multi-antenna systems and cannot be directly applied to PA systems.

%

To fill this gap, we propose a differential spatial modulation scheme tailored for PA systems, termed as {\em DSM-PA}.
By embedding both the Alamouti coding structure and pinching position IM into a unified differential codeword, the proposed scheme enables noncoherent transmission without requiring instantaneous CSI while achieving full transmit diversity. Furthermore, it effectively exploits the spatial degrees of freedom of PA systems to enhance spectral efficiency. The main contributions of this paper are summarized as follows:

\begin{itemize}

\item
We propose a novel DSM-based transmission scheme for PA systems to eliminate the reliance on instantaneous CSI at the receiver.
By integrating the Alamouti coding structure and pinching-position IM into the differential codeword, the proposed design simultaneously enables noncoherent transmission, achieves full transmit diversity, and enhances spectral efficiency.

\item
We derive an asymptotically tight upper bound on the average bit error rate (BER) by using the moment-generating function (MGF) approach. We also verify that the proposed DSM-PA scheme can achieve full transmit diversity, which is determined by the rank of the codeword difference matrix.

\item
Simulation results validate the tightness of the derived BER upper bounds in the high signal-to-noise ratio (SNR). Our results also demonstrate that the proposed DSM-PA scheme outperforms the benchmark, while maintaining a reasonable BER loss with respect to the coherent scheme with perfect CSI.

\end{itemize}

{\em Notations:}
The operators $(\cdot)^{\rm T}$, $(\cdot)^{\rm H}$, and $(\cdot)^{-1}$ denote the transpose, conjugate transpose, and matrix inversion, respectively.
The symbol $\lfloor \cdot \rfloor$ denotes the floor operation, and $\|\cdot\|$ denotes the Euclidean norm.
The notations $\operatorname{diag}(\cdot)$ and $\operatorname{blkdiag}(\cdot)$ denote a diagonal matrix and a block-diagonal matrix formed by their arguments, respectively.
The notations $\mathbb{C}^{M\times N}$ and $\mathbb{R}^{M\times N}$ are the sets of $M\times N$ complex-valued and real-valued matrices, respectively, and $\mathbf{I}_N$ denotes the $N\times N$ identity matrix.
The notation $\mathcal{CN}(\boldsymbol{\mu},\mathbf{\Sigma})$ represents a complex Gaussian distribution with mean vector $\boldsymbol{\mu}$ and covariance matrix $\mathbf{\Sigma}$.
The operator $\mathbb{E}\{\cdot\}$ denotes expectation, and $\mathbb{E}\{\cdot\mid\cdot\}$ denotes conditional expectation.

\begin{figure}[t]
	\center
	\includegraphics[width=3.3in,height=1.3in]{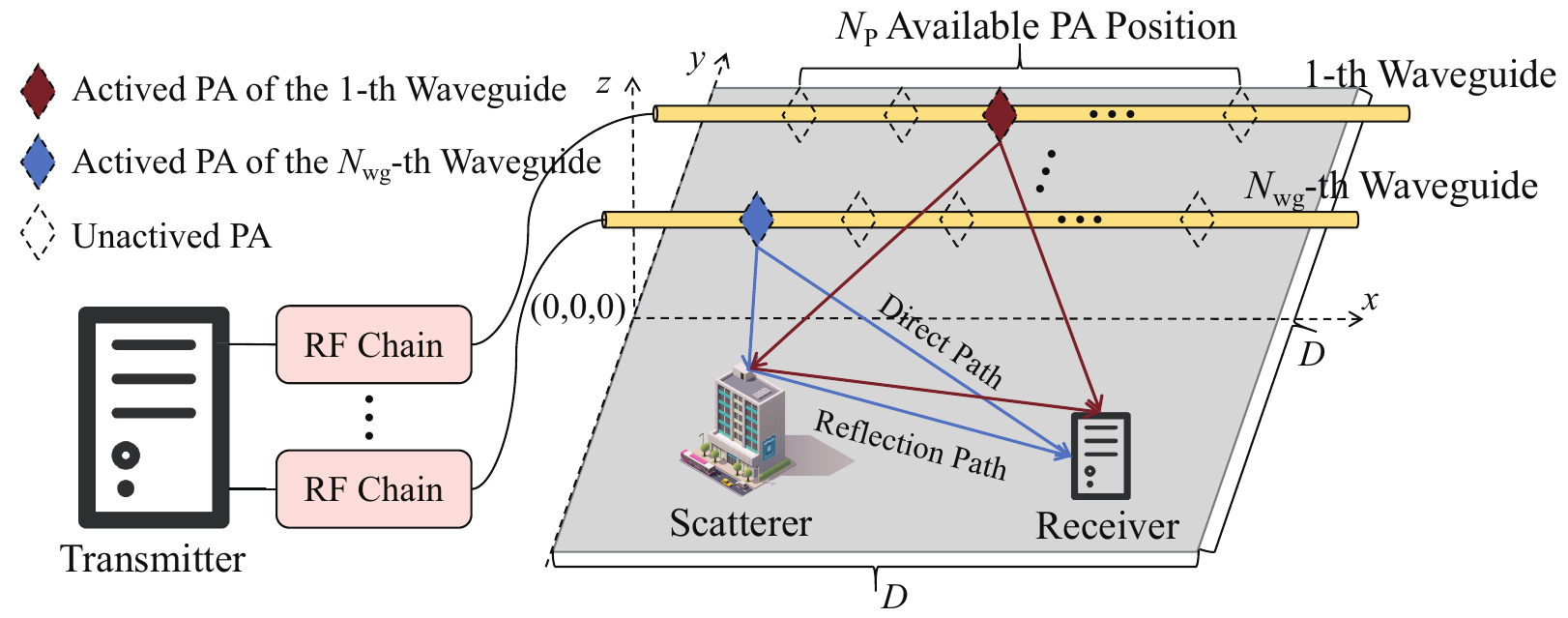}
	\vspace{-0.2cm}
	\caption{{A schematic of the proposed DSM-PA system for a downlink single-user scenario.}}
	\label{system-model}  
	\vspace{-3mm}
\end{figure}
\begin{figure}[t]
	\center
	\includegraphics[width=3.2in,height=1.3in]{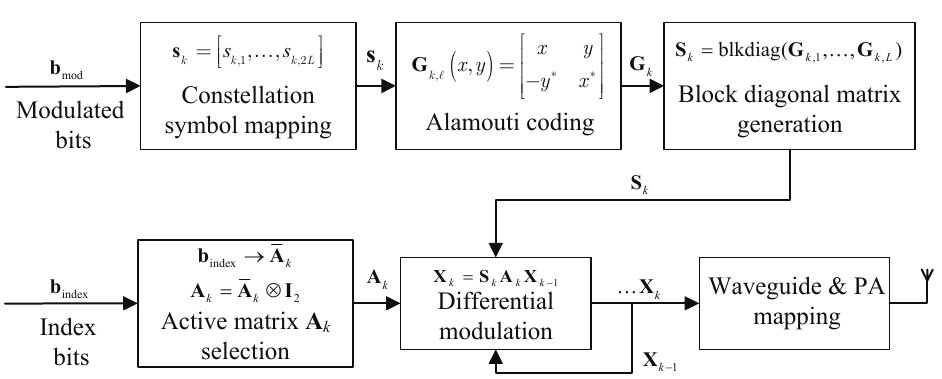}
	\vspace{-0.2cm}
	\caption{{Block diagram of the transmitter for the proposed DSM-PA scheme.}}
	\label{transmitter}  
	\vspace{-3mm}
\end{figure}
\section{Principle of the Proposed DSM-PA Scheme}
As shown in Fig.~\ref{system-model}, we present a schematic diagram of the proposed DSM-PA system for a downlink single-user scenario, where the transmitter is equipped ${N}_{\rm wg}$ waveguides operating within a region of side length $D$.
Each waveguide has $N_{\rm p}$ available pinching positions and is equipped with a dedicated RF chain, thereby enabling simultaneous and independent information transmission within the same time slot.

\subsection{Transmitter of the Proposed DSM-PA Scheme}
With loss of generality, we consider a dual-waveguide PA structure,\footnote{This paper focuses on the dual-waveguide configuration, i.e., $N_{\mathrm{wg}}=2$, which represents the minimum structure required to construct an Alamouti subblock. This setting is sufficient to highlight the core design principles of the proposed DSM-PA scheme. More general multi-waveguide scenarios can be extended via waveguide grouping or higher-dimensional block designs. However, such extensions introduce additional notation and implementation complexity without altering the fundamental conclusions on noncoherent design and full diversity.} where radiating elements from the two waveguides at each pinching position form an Alamouti transmit pair to achieve full transmit diversity.
Specifically, Fig.~\ref{transmitter} illustrates the transmitter structure of the proposed DSM-PA scheme, where the input bits are divided into modulation bits ${\bf b}_{\rm mod}$ and index bits ${\bf b}_{\rm index}$.
Taking the $k$-th DSM-PA transmitted block as an example, the modulation bits are mapped to $\mathcal{M}$-phase shift keying (PSK) to form the modulated  symbol vector $\mathbf{s}_{k}=[s_{k,1},s_{k,2},\ldots,s_{k,T}]^{\mathrm{T}}\in{\mathbb{C}^{T\times 1}}$, where $k\in\{0,1,\ldots\}$, $\mathcal{M}$ is the modulation order, and $T$ represents the number of time slots for each transmitted block. Hence, the number of modulated bits can be calculated as $B_{\rm mod}=T{{\rm log}_2}{\mathcal{M}}$.
Next, the modulated symbol vector is divided into pairs, each forming an Alamouti subblock, and the $\ell$-th subblock ${\mathbf{G}}_{k,\ell}\in{\mathbb{C}^{2\times2}}$ is defined as
\begin{align}\label{eq3}
{\mathbf{G}}_{k,\ell}\left(s_{k,2\ell-1},s_{k,2\ell}\right)
= \frac{1}{\sqrt{2}}
\begin{bmatrix}
s_{k,2\ell-1} & s_{k,2\ell} \\
- s_{k,2\ell}^{*} & s_{k,2\ell-1}^{*}
\end{bmatrix},
\end{align}
where $\ell =\{ 1,2,\ldots,L\}$, and we define $L=N_{\rm p}=T/2$ denotes the number of Alamouti subblocks.
The $L$ Alamouti subblocks are combined to form a block-diagonal modulation matrix ${\mathbf{S}}_{k}\in{\mathbb{C}^{T\times T}}$, which can be given by
\begin{align}\label{eq4}
{\mathbf{S}}_{k}={\rm blkdiag}\{{\mathbf{G}}_{k,1},\ldots,{\mathbf{G}}_{k,\ell},\ldots,{\mathbf{G}}_{k,L}\}.
\end{align}
In the modulation matrix ${\mathbf{S}}_{k}$, the column index $c$ corresponds to the time slot, whereas the row index $r=(\ell-1){N_{\rm wg}}+n$ represents the joint index of the waveguide and pinching position. In other words, the symbol $S_k[r,c]$ corresponds to the transmission at time slot $c$ from the $\ell$-th pinching position of the $n$-th waveguide, where $n =\{ 1,2\}$.

%

It can be observed from~\eqref{eq4} that the mapping between the transmitted symbols and the corresponding PA pinching position remains fixed, which fails to exploit the spatial degrees of freedom offered by multiple pinching positions in PA systems. To address this limitation, we incorporated IM technique in the pinching-position domain to further enhance the spectral efficiency of the DSM-PA system.
Specifically, the index bits ${\bf b}_{\rm index}$ are used to determine the assignment of the $L$ Alamouti subblocks over the $N_{\rm p}$ pinching positions. Since the two radiating elements generated from the two waveguides at the same pinching position form one Alamouti transmission pair, the index mapping is performed on the Alamouti-subblock level. Let
\begin{align}\label{eq5}
\mathcal{A}=\{\bar{\mathbf{A}}_{1},\bar{\mathbf{A}}_{2},\ldots,\bar{\mathbf{A}}_{2^{B_{\mathrm{index}}}}\}
\end{align}
denote the IM codebook, where $\forall \bar{\mathbf{A}}_{u}\in\mathbb{R}^{L\times L}$ is a permutation matrix and ${B_{\mathrm{index}}}={{\mathrm{log}}_2\lfloor L! \rfloor}$ is the number of index bits.
For the $k$-th DSM-PA transmitted block, the index bits ${\bf b}_{\mathrm{index}}$ are mapped to a codeword $\bar{\mathbf{A}}_{u,k}$, which is selected from the IM codebook $\mathcal{A}$. In this way, the selected codeword specifies a one-to-one assignment of the $L$ Alamouti subblocks to the ${N_{\rm p}}$ pinching positions. For example, when \(L=4\), if the first, second, third, and fourth Alamouti subblocks are assigned to the third, first, fourth, and second pinching positions, respectively, the selected permutation matrix of $k$-th differential block is given by
\begin{align}\label{eq6}
\bar{\mathbf{A}}_{u_k}=
\begin{bmatrix}
0 & 0 & 1 & 0\\
1 & 0 & 0 & 0\\
0 & 0 & 0 & 1\\
0 & 1 & 0 & 0
\end{bmatrix}.
\end{align}
Accordingly, the activation matrix in the \(T\)-dimensional signal space is given by $\mathbf{A}_{k}=\bar{\mathbf{A}}_{u_k}\otimes\mathbf{I}_{2}\in\mathbb{R}^{T\times T}$,
where $\otimes$ denotes the Kronecker product. The matrix \(\mathbf{A}_{k}\) rearranges the $L$ Alamouti subblocks according to the selected index pattern, while keeping the internal $2\times 2$ structure of each subblock unchanged.
Then, by combining the modulation matrix \(\mathbf{S}_{k}\) and the activation matrix \(\mathbf{A}_{k}\), the $k$-th differential codeword can be written as $\mathbf{Q}_{k}=\mathbf{S}_{k}\mathbf{A}_{k}\in\mathbb{C}^{T\times T}$.
Finally, the $k$-th transmitted symbol matrix $\mathbf{X}_{k}\in\mathbb{C}^{T\times T}$ is recursively generated as
\begin{align}\label{eq9}
\mathbf{X}_{k}=\mathbf{Q}_{k}\mathbf{X}_{k-1},
\end{align}
where $\mathbf{X}_{k-1}$ denotes the $(k-1)$-th transmitted matrix, and ${\bf X}_{0}={\bf I}_{T}$ is the initial reference matrix. 
This differential encoding structure conveys information through the relative variation between consecutive transmission blocks, thereby enabling noncoherent detection and eliminating the need for instantaneous CSI at the receiver, which significantly reduces pilot overhead.

\subsection{Channel Model}
We consider a quasi-static Rician fading channel, where the channel coefficients remain constant over the entire transmission frame, thereby ensuring that adjacent transmission blocks experience the same channel realization.
Based on the transmitted matrix in~\eqref{eq9}, the equivalent channel can be represented as a vector ${\bf h}\in{\mathbb{C}}^{T\times 1}$ given by
\begin{align}\label{eq10}
\mathbf{h}=[h_{1,1},h_{2,1},\ldots,h_{1,\ell},h_{2,\ell},\ldots,h_{{1},{N_{\mathrm{p}}}},h_{{2},{N_{\mathrm{p}}}}]^{\rm T}.
\end{align}
Let $h_{n,\ell}$ denote the equivalent channel coefficient from the $n$-th waveguide at the $\ell$-th pinching position to the receiver, which can be given by~\cite{11368709}
\begin{align}\label{channel}
h_{n,\ell}\! =\! \sqrt{\beta_{n,\ell}} \!\left( 
\!\sqrt{\frac{K_{n,\ell}}{K_{n,\ell}+1}} \, \bar{h}_{n,\ell}
\!+ \!\sqrt{\frac{1}{K_{n,\ell}+1}} \, \tilde{h}_{n,\ell}\!
\right)\! \gamma_{n,\ell},
\end{align}
where $\beta_{n,\ell}$ is the large-scale fading coefficient, $K_{n,\ell}$ is the Rician factor, $\bar{h}_{n,\ell}$ is the deterministic line-of-sight (LoS) component, and $\tilde{h}_{n,\ell} \sim \mathcal{CN}(0,1)$ is the scattered component.
Moreover, $\gamma_{n,\ell}$ denotes the position-dependent phase shift introduced by the $\ell$-th pinching position. Specifically, letting $\mathbf{p}_{n,\ell}$ and $\mathbf{p}_{r}$ denote the coordinates of the $\ell$-th pinching position on the $n$-th waveguide and the receiver, respectively, the corresponding propagation distance is given by
\begin{align}\label{channel.1}
d_{n,\ell}=\|\mathbf{p}_{r}-\mathbf{p}_{n,\ell}\|.
\end{align}
Accordingly, the large-scale fading coefficient and the position-dependent phase term are respectively given by $\beta_{n,\ell}=\left(\frac{1}{d_{n,\ell}}\right)^{\alpha}$ and $\gamma_{n,\ell}=\exp\!\left(-j\frac{2\pi}{\lambda_c}d_{n,\ell}\right)$, where $\alpha$ denotes the path-loss exponent and $\lambda_c$ denotes the carrier wavelength.
Furthermore, $\mathbf h$ follows a complex Gaussian distribution, i.e., $\mathbf h\sim\mathcal{CN}(\boldsymbol{\mu}_h,\mathbf\Sigma_h)$, where $\boldsymbol{\mu}_h$ follows the same ordering as $\mathbf h$ in~\eqref{eq10}, with $(n,\ell)$-th entry $\mu_{n,\ell}=\sqrt{\beta_{n,\ell}}\sqrt{\frac{K_{n,\ell}}{K_{n,\ell}+1}}\,\bar h_{n,\ell}\gamma_{n,\ell}$. 
$\mathbf\Sigma_h$ is a diagonal covariance matrix whose $(n,\ell)$-th diagonal entry is $\sigma_{n,\ell}^{2}=\frac{\beta_{n,\ell}}{K_{n,\ell}+1}$. Since $\gamma_{n,\ell}$ is a unit-modulus phase factor, it only affects the phase of $\mu_{n,\ell}$ and does not change the corresponding variance $\sigma_{n,\ell}^{2}$.

{\bf{\em Remark 1:}} According to \eqref{channel}, different pinching positions generally yield different channel coefficients. Hence, coherent detection requires the receiver to acquire the instantaneous CSI of all $h_{n,\ell}$. As $N_p$ increases, the number of position-dependent channels to be estimated and tracked grows rapidly, resulting in significant pilot overhead and CSI acquisition complexity. Therefore, noncoherent differential detection is adopted in the proposed DSM-PA scheme.

\subsection{Receiver of the Proposed DSM-PA Scheme}
Next, the received signal of the $k$-th transmitted block can then be written as
\begin{align}\label{eq15}
\mathbf y_k=\mathbf X_k\mathbf h+\mathbf n_k,
\end{align}
where $\mathbf n_k\sim\mathcal{CN}(\mathbf 0,N_0\mathbf I_T)\in\mathbb{C}^{T\times 1}$ denotes additive white Gaussian noise of the $k$-th received signal. By using the differential encoding rule in \eqref{eq9}, \eqref{eq15} can be rewritten as
\begin{align}\label{eq16}
& {{\mathbf{y}}_{k}}=\underbrace{{{\mathbf{Q}}_{k}}{{\mathbf{X}}_{k-1}}}_{{{\mathbf{X}}_{k}}}\mathbf{h}+{{\mathbf{n}}_{k}} \nonumber\\ 
 & ={{\mathbf{Q}}_{k}}\underbrace{\left( {{\mathbf{X}}_{k-1}}\mathbf{h}+{{\mathbf{n}}_{k-1}} \right)}_{{{\mathbf{y}}_{k-1}}}+\underbrace{{{\mathbf{n}}_{k}}-{{\mathbf{Q}}_{k}}{{\mathbf{n}}_{k-1}}}_{{\tilde {\bf n}}_{k}} \nonumber\\ 
 & ={{\mathbf{Q}}_{k}}{{\mathbf{y}}_{k-1}}+{\tilde {\bf n}}_{k},
\end{align}
where ${{\bf{y}}_{k-1}}$ represents the $(k-1)$-th received signal, and ${\tilde {\bf n}}_{k}\triangleq{{\bf n}}_{k}-{\bf Q}_{k}{{\bf n}}_{k-1}$ denotes the equivalent noise. Since ${\bf Q}_{k}$ is a unitary matrix and ${{\bf n}}_{k}$ and ${{\bf n}}_{k-1}$ are independent of each other,  we have ${\tilde{\mathbf n}}_k\sim\mathcal{CN}(\mathbf 0,2N_0\mathbf I_T)$.
It follows from \eqref{eq16} that $\mathbf y_{k-1}$ serves as a noisy observation of the effective channel $\mathbf X_{k-1}\mathbf h$. Therefore, the transmitted matrix $\mathbf Q_k$ can be detected by jointly processing two adjacent received blocks, i.e., $\mathbf y_k$ and $\mathbf y_{k-1}$, without explicitly estimating the instantaneous CSI. In this way, the proposed DSM-PA scheme enables noncoherent differential detection.
Accordingly, differential maximum-likelihood (ML) detection is adopted at the receiver. Specifically, given the current observation $\mathbf y_k$ and the previous received block $\mathbf y_{k-1}$, the transmitted codeword is detected as
\begin{align}\label{eq17}
\hat{\mathbf Q}_k
=
\arg\min_{\mathbf Q_{i}\in\mathcal Q}
\left\|
\mathbf y_k-\mathbf Q_{i}\mathbf y_{k-1}
\right\|^2,
\end{align}
where $\mathcal Q$ denotes the set of all legitimate transmit matrices with $\mathbf Q_k=\mathbf S_k\mathbf A_k$. Once $\hat{\mathbf Q}_k$ is obtained, the corresponding activation pattern and modulation symbols are demapped to recover the index bits and modulated bits, respectively.

\section{Performance Analysis}
In this section, we analyze the BER performance and diversity gain of the proposed DSM-PA scheme. Specifically, the pairwise error probability (PEP) is first derived based on the differential ML detection criterion. Then, a high SNR asymptotic analysis is performed, demonstrating that the proposed DSM-PA scheme can achieve full transmit diversity.
\begin{figure*}[t]
\begin{align}\label{eq23}
\mathcal{M}_{Z_{ij}\mid \mathbf{X}_{k-1}}(s)
=
\frac{
\exp\!\left(
-s\,
\boldsymbol{\mu}_{g,k-1}^{H}
\mathbf{A}_{ij}
\left(
\mathbf{I}_{T}+s\mathbf{\Sigma}_{g,k-1}\mathbf{A}_{ij}
\right)^{-1}
\boldsymbol{\mu}_{g,k-1}
\right)
}{
\det\!\left(
\mathbf{I}_{T}+s\mathbf{\Sigma}_{g,k-1}\mathbf{A}_{ij}
\right)
}.
\tag{16}
\end{align}
\noindent\rule{\textwidth}{0.5pt}
\end{figure*}
\begin{figure*}[t]
\begin{align}\label{eq24}
{\overline{\rm Pr}}_{ij\mid \mathbf{X}_{k-1}}
=
\frac{1}{\pi}
\int_{0}^{\pi/2}
\frac{
\exp\!\left(
-\xi(\theta)\,
\boldsymbol{\mu}_{g,k-1}^{H}
\mathbf{A}_{ij}
\left(
\mathbf{I}_{T}+\xi(\theta)\mathbf{\Sigma}_{g,k-1}\mathbf{A}_{ij}
\right)^{-1}
\boldsymbol{\mu}_{g,k-1}
\right)
}{
\det\!\left(
\mathbf{I}_{T}+\xi(\theta)\mathbf{\Sigma}_{g,k-1}\mathbf{A}_{ij}
\right)
}
\,d\theta.
\tag{17}
\end{align}
\noindent\rule{\textwidth}{0.5pt}
\end{figure*}
\subsection{MGF-based Semi-analytical BER Upper Bound}
We consider the event that the transmitted symbol ${\bf Q}_{i}$ is erroneously detected as ${\bf Q}_{j}$. The corresponding PEP can be written as
\begin{align}\label{eq19}
{\rm Pr}(\mathbf Q_i\!\rightarrow \!\mathbf Q_j)
\!=\!
\Pr\!\left(\!
\left\|
\mathbf y_k\!-\!\mathbf Q_j\mathbf y_{k-1}
\right\|^2
\!\le\!
\left\|
\mathbf y_k\!-\!\mathbf Q_i\mathbf y_{k-1}
\right\|^2\!
\right).
\end{align}
It is worth noting that both the decision statistic $\mathbf{y}_{k-1}$ and the equivalent noise ${\tilde {\bf n}}_{k}$ contain the noise component $\mathbf{n}_{k-1}$. Therefore, deriving the exact error probability would require handling correlated noise terms, which makes it difficult to obtain a concise closed-form BER expression. To overcome this issue, we adopt the widely used high-SNR approximation in noncoherent differential detection, where the decision statistic $\mathbf{y}_{k-1}$ is approximated by the noiseless effective channel vector $\mathbf g_{k-1}\triangleq \mathbf X_{k-1}\mathbf h$. 
Under this approximation, we define the difference matrix as ${\bf\Delta}_{ij} = {\bf Q}_i - {\bf Q}_j$, and ${\bf A}_{ij} = {\bf\Delta}_{ij}^H {\bf\Delta}_{ij}$. Hence,  the conditional PEP can be given by
\begin{align}\label{eq20}
{\rm Pr}({\bf Q}_i \to {\bf Q}_j \mid {\bf g}_{k-1}) \approx Q\left(\sqrt{\frac{{\bf g}_{k-1}^H {\bf A}_{ij} {\bf g}_{k-1}}{4 N_0}}\right).
\end{align}
For a given previous differential state ${\bf X}_{k-1}$, the vector $\mathbf g_{k-1}\triangleq \mathbf X_{k-1}\mathbf h$ remains random due to the randomness of the channel ${\bf h}$. Therefore, the PEP averaged over the channel realization can be obtained by taking the expectation of \eqref{eq20} conditioned on ${\bf X}_{k-1}$. To this end, define the random variable ${Z}_{ij}\triangleq {\bf g}_{k-1}^{H}{\bf A}_{ij}{\bf g}_{k-1}$, the channel-averaged PEP can be expressed as
\begin{align}\label{eq21}
{\overline{\rm Pr}}_{ij\mid \mathbf{X}_{k-1}}
=
\int_{0}^{\infty}
Q\!\left(
\sqrt{\frac{z}{4N_0}}
\right)
p_{{Z}_{ij}\mid \mathbf{X}_{k-1}}(z)\,dz,
\end{align}
where $p_{{Z}_{ij}\mid \mathbf{X}_{k-1}}(z)$ denotes the conditional probability density function (PDF) of ${Z}_{ij}$.
Next, by invoking Craig's formula  $Q(x)=\frac{1}{\pi}\int_{0}^{\pi/2}
\exp\!\left(
-\frac{x^2}{2\sin^2\theta}
\right)d\theta$ for the Gaussian $Q$-function, \eqref{eq21} can be reformulated as
\begin{align}\label{eq22}
{\overline{\rm Pr}}_{ij\mid \mathbf{X}_{k-1}}
=
\frac{1}{\pi}
\int_{0}^{\pi/2}
\mathcal{M}_{Z_{ij}\mid \mathbf{X}_{k-1}}
\!\left(
\xi(\theta)
\right)
d\theta,
\end{align}
where $\xi(\theta)\triangleq \frac{1}{8N_0\sin^2\theta}$, and $
\mathcal{M}_{Z_{ij}\mid \mathbf{X}_{k-1}}(s)
\triangleq
\mathbb{E}
\!\left[
e^{-sZ_{ij}}
\,\big|\,
\mathbf{X}_{k-1}
\right]$
is the conditional MGF of $Z_{ij}$. 
Since $\mathbf{h}\sim\mathcal{CN}(\boldsymbol{\mu}_h,\mathbf{\Sigma}_h)$ and $\mathbf{g}_{k-1}=\mathbf{X}_{k-1}\mathbf{h}$, we can obtain
$\mathbf{g}_{k-1}\sim
\mathcal{CN}\!\left(
\boldsymbol{\mu}_{g,k-1},
\mathbf{\Sigma}_{g,k-1}
\right)$. As $\mathbf{X}_{k-1}$ is a unitary matrix, we can calculate the mean and variance as $\boldsymbol{\mu}_{g,k-1}=\mathbf{X}_{k-1}\boldsymbol{\mu}_h$ and $\mathbf{\Sigma}_{g,k-1}=\mathbf{X}_{k-1}\mathbf{\Sigma}_h\mathbf{X}_{k-1}^{H}$, respectively.
Next, using the standard result for the MGF~\cite{11368709} of a noncentral Hermitian quadratic form in a complex Gaussian vector, the conditional MGF can be expressed as~\eqref{eq23}. Substituting \eqref{eq23} into \eqref{eq22}, the conditional averaged PEP is obtained as~\eqref{eq24}.

Finally, averaging \eqref{eq24} over the previous differential state $\mathbf{X}_{k-1}$, the unconditional PEP is given by
\begin{align}\label{eq25}
{\rm Pr}\!\left(\mathbf{Q}_i \rightarrow \mathbf{Q}_j\right)
=
\mathbb{E}_{\mathbf{X}_{k-1}}
\!\left[
{\overline{\rm Pr}}_{ij\mid \mathbf{X}_{k-1}}
\right].
\tag{18}
\end{align}
Since the previous differential state $\mathbf{X}_{k-1}$ depends recursively on the entire sequence of past transmitted codewords, the outer expectation in \eqref{eq25} is generally intractable to evaluate in closed form. To this end, \eqref{eq25} is regarded as a semi-analytical PEP. Also, the one-dimensional integral in \eqref{eq24} can be computed efficiently using standard numerical quadrature method, while the remaining expectation over $\mathbf{X}_{k-1}$ is evaluated numerically via sample averaging.
\setcounter{equation}{18}

Finally, based on the obtained PEP in~\eqref{eq25}, we can calculate the MGF-based semi-analytical BER upper bound for the proposed DSM-PA scheme~as
\begin{align}\label{eq26}
{\rm Pr}_{\rm BER} \le 
\frac{1}{2^{B}B}
\sum_{\mathbf{Q}_i}
\sum_{\mathbf{Q}_j}
N(\mathbf{b}_i,\mathbf{b}_j)\,
{\rm Pr}\!\left(\mathbf{Q}_i \rightarrow \mathbf{Q}_j\right),
\end{align}
where $B=B_{\rm mod}+B_{\rm index}$ denotes the number of total information bits carried by the $k$-th differential block, and  $N(\mathbf{b}_i,\mathbf{b}_j)$ denotes the number of bit errors when $\mathbf{Q}_i$ is detected as $\mathbf{Q}_j$.

\subsection{Diversity Analysis}
Based on the MGF-based PEP expression in \eqref{eq24}, the diversity gain of the proposed DSM-PA scheme can be characterized through its high-SNR asymptotics. 
Let us define $\mathbf{B}_{ij} \triangleq\mathbf{\Sigma}_{g,k-1}^{1/2}\mathbf{A}_{ij}\mathbf{\Sigma}_{g,k-1}^{1/2}$, $r_{ij}$ denote the rank of $\mathbf{B}_{ij}$, and ${\bm \lambda}=[{\lambda}_{1},\ldots,{\lambda}_{\phi},\ldots{\lambda}_{r_{ij}}]$ denote the nonzero eigenvalues vector of $\mathbf{B}_{ij}$. Since $\mathbf{B}_{ij}$ is Hermitian positive semidefinite, the determinant term in \eqref{eq24} can be expressed as
\begin{align}\label{eq27}
\det\!\left(
\mathbf{I}_T+\xi(\theta)\mathbf{\Sigma}_{g,k-1}\mathbf{A}_{ij}
\right)
&=
\det\!\left(
\mathbf{I}_T+\xi(\theta)\mathbf{B}_{ij,k-1}
\right)\nonumber\\
&=
\prod_{\phi=1}^{r_{ij}}
\left(
1+\xi(\theta)\lambda_{\phi}
\right).
\end{align}
In the high-SNR region, one can find that $\xi(\theta)\to \infty$, which yields
\begin{equation}
\prod_{\phi=1}^{r_{ij}}
\left(
1+\xi(\theta)\lambda_{\phi}
\right)
\sim
\xi(\theta)^{r_{ij}}
\prod_{\ell=1}^{r_{ij}}\lambda_{\phi}.
\label{eq:det_asym}
\end{equation}
Meanwhile, the exponential term in \eqref{eq24} remains bounded and therefore affects only the coding gain rather than the asymptotic decay order. Consequently, the conditional average PEP satisfies ${\overline{\rm Pr}}_{ij\mid \mathbf{X}_{k-1}}\doteq\gamma^{-r_{ij}}$, where $\doteq$ denotes exponential equality, and $\gamma={E_b}/{N_0}$. It follows that the diversity of the proposed DSM-PA scheme is given by
\begin{equation}
d
=
-\lim_{E_b/N_0\to\infty}
\frac{\log {\rm Pr}_{\rm BER}}{\log(E_b/N_0)}
=
\min_{i\neq j} r_{ij}.
\label{eq:diversity_def}
\end{equation}
Because $\mathbf{\Sigma}_{g,k-1}$ is positive definite, $\mathbf{\Sigma}_{g,k-1}^{1/2}$ is nonsingular. Therefore, $\operatorname{rank}(\mathbf{B}_{ij})=\operatorname{rank}(\mathbf{A}_{ij})$. Moreover, since $\mathbf{A}_{ij}=\mathbf{\Delta}_{ij}^{H}\mathbf{\Delta}_{ij}$, it follows that $\operatorname{rank}(\mathbf{A}_{ij})=\operatorname{rank}(\mathbf{\Delta}_{ij})$. Hence,
\begin{equation}
d=\min_{i\neq j}\operatorname{rank}(\mathbf{B}_{ij})
=\min_{i\neq j}\operatorname{rank}(\mathbf{\Delta}_{ij}).
\label{eq:div_rank_delta_compact}
\end{equation}
According to Section II, we can see that ${\bf Q}_{k}$ consists of $L$ independent Alamouti subblocks. 
Thus, any nonzero difference matrix $\mathbf{\Delta}_{ij}$ contains at least one nonzero Alamouti subblock, whose determinant is $(|s_{k,2\ell-1}|^2+|s_{k,2\ell}|^2)/2>0$. Therefore, the rank of the difference matrix $\mathbf{\Delta}_{ij}$ is greater than or equal to 2, i.e., $\operatorname{rank}(\mathbf{\Delta}_{ij})\ge 2$. Hence, the diversity of the proposed DSM-PA scheme is given by $d=\min_{i\neq j}\operatorname{rank}(\mathbf{\Delta}_{ij})=2$.
This result indicates that the proposed DSM-PA architecture can achieve full transmit diversity, in which the index bits determine the activated pinching-position pair across the two waveguides, and the modulation symbols are jointly encoded within the corresponding Alamouti block. As a result, the minimum nonzero error event is associated with a single active $2\times2$ Alamouti block, thereby yielding a diversity order of two.

\begin{figure}[t]
	\center
	\includegraphics[width=2.3in,height=1.6in]{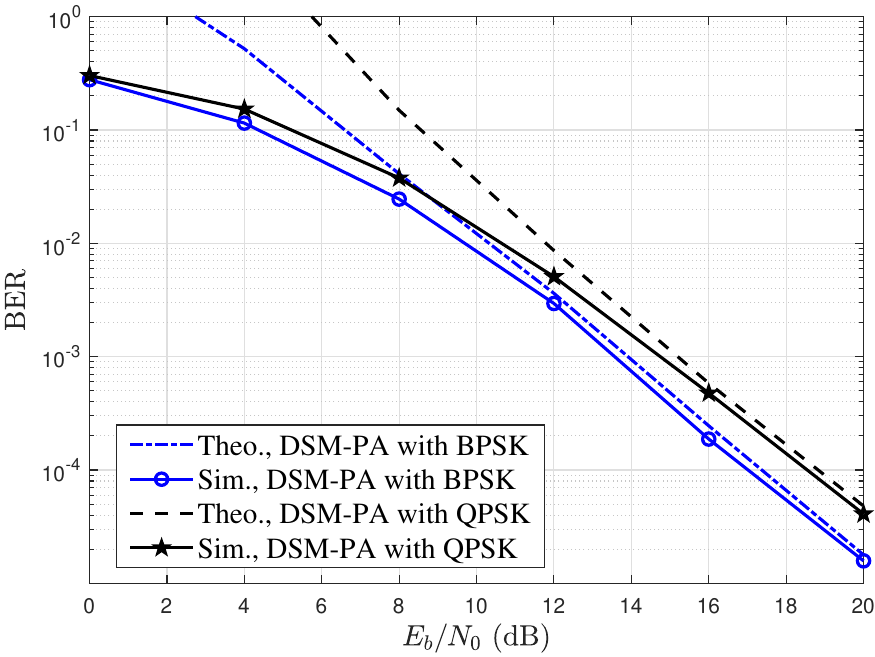}
	\vspace{-0.2cm}
	\caption{{Comparison between the theoretical BER upper bound and the simulated BER of the proposed DSM-PA scheme for BPSK and QPSK.}}
	\label{Theory_BER}  
	\vspace{-4mm}
\end{figure}
\section{Simulation Results and Discussions}
In this section, the performance of the proposed DSM-PA scheme is evaluated. We consider a downlink single-user transmission scenario, where the proposed DSM-PA scheme operates within a square region of side length of $D=20~{\rm m}$, and the carrier frequency is set to $f_c=28~{\rm GHz}$.  The transmitter is equipped with two waveguides deployed in parallel along the $x$-axis, with a waveguide height of $3~{\rm m}$ and the inter-waveguide spacing of $0.5~{\rm m}$. The receiver is fixed at $\mathbf{p}_{r}=[13.2,~7.1,~0]~{\rm m}$. The channel follows the quasi-static Rician fading model described in Section II, where the path-loss exponent is set to $\alpha=2.2$, and the Rician factor is set as $5$. Unless otherwise stated, the number of waveguides is set to $N_{\rm wg}=2$, the number of pinching positions on each waveguide is set to $N_{\rm p}=3$, and BPSK modulation is adopted. A fixed nonuniform pinching-position deployment is adopted, where the position set along each waveguide is given by $\mathcal{X}_{\rm position}=\{1.5,\ 7.5,\ 18.5\}~{\rm m}$.
Hence, the coordinate of the \(\ell\)-th pinching position on the $n$-th waveguide is $\mathbf{p}_{n,\ell}=[x_\ell,\ y_n,\ 3]~{\rm m}$.

First, we compare the theoretical BER curves with the simulation results for the proposed DSM-PA scheme under BPSK and QPSK in~Fig.~\ref{Theory_BER}. It can be observed that the theoretical results closely match the simulated ones over the high-SNR region, which validates the effectiveness of the MGF-based BER analysis developed in Section III-A. In addition, the BPSK-based DSM-PA scheme achieves better BER performance than its QPSK counterpart, especially in the high-SNR region. This can be explained by the larger minimum Euclidean distance of the BPSK constellation.
Besides, a slight gap between the theoretical and simulated curves may be observed at low SNRs, which is mainly caused by the approximation adopted in the differential detection derivation, where the decision statistic $\mathbf{y}_{k-1}$ is replaced by the noiseless effective channel vector $\mathbf{g}_{k-1}$. This approximation becomes less accurate in the low-SNR regime due to the stronger noise effect.

\begin{figure}[t]
        \centering
        \includegraphics[width=2.3in,height=1.6in]{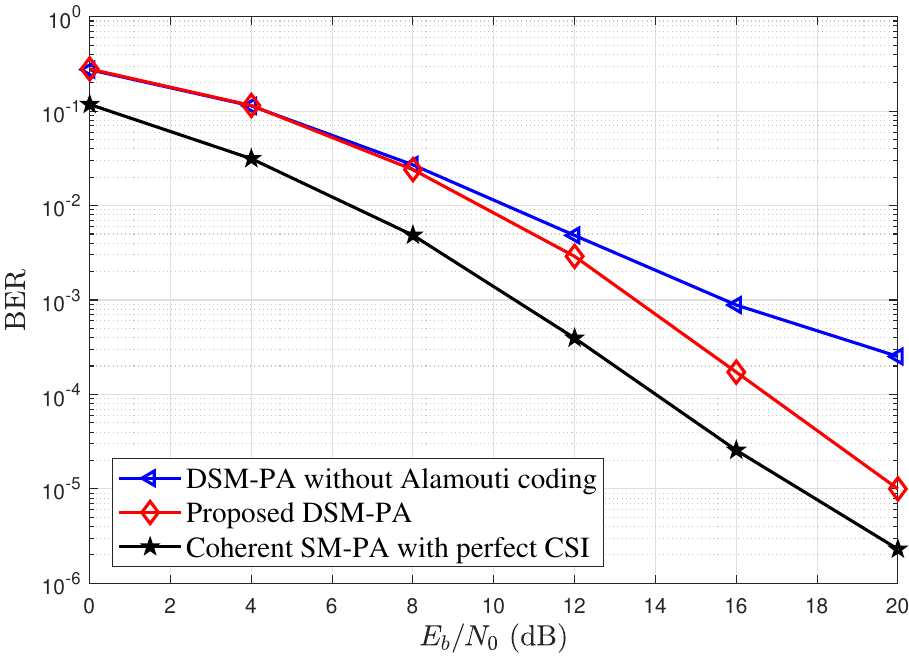}
        \vspace{-0.2cm}
    	\caption{ BER comparison of the proposed DSM-PA, the coherent SM-PA scheme with perfect CSI, and the DSM-PA schemes without Alamouti coding.
    }
    	\label{SIM4}  
    	\vspace{-3mm}
\end{figure}

Fig.~\ref{SIM4} compares the BER performance of the proposed DSM-PA scheme with two benchmark schemes, i.e., coherent SM-PA with perfect CSI and DSM-PA scheme without Alamouti coding. It can be observed that the proposed DSM-PA scheme incurs an SNR loss of approximately $3~{\rm dB}$ with respect to the coherent SM-PA with perfect CSI scheme at the $10^{-3}$ BER level. This gap is reasonable, since the proposed scheme avoids instantaneous CSI acquisition and differential noncoherent detection generally suffers from the well-known SNR penalty relative to coherent detection. 
On the other hand, compared to the DSM-PA scheme without Alamouti coding, the proposed scheme exhibits a markedly steeper BER decay in the high-SNR regime, experiencing a higher diversity order. This indicates that the introduced Alamouti coding structure effectively provides transmit diversity. These observations are consistent with the diversity analysis in Section III-B.

\begin{figure}[t]
	   \centering
        \includegraphics[width=2.3in,height=1.6in]{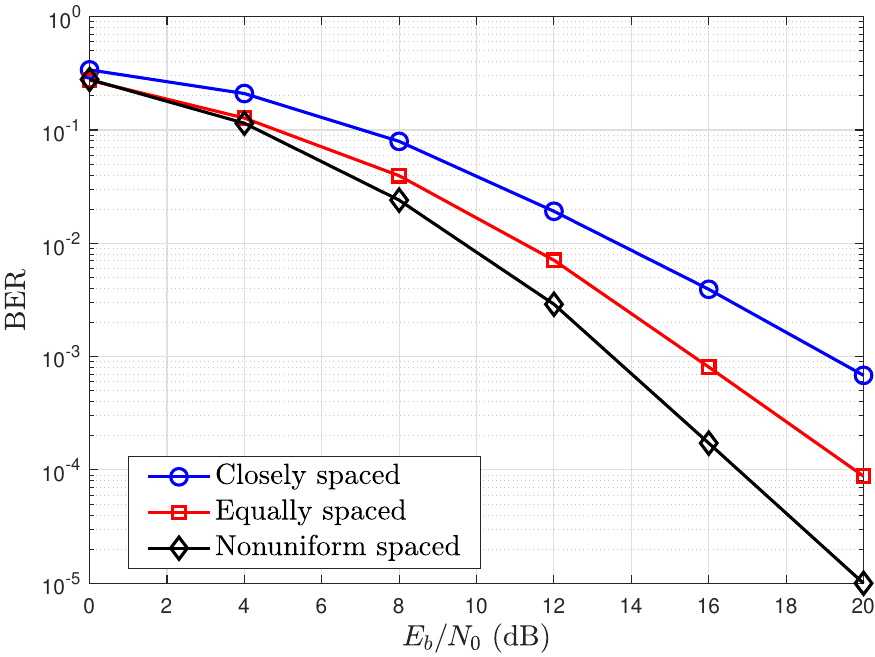}
        \vspace{-0.2cm}
	\caption{{BER comparison of the proposed DSM-PA scheme under different pinching-position deployments.}}
	\label{deployment_PA}  
	\vspace{-5mm}
\end{figure}
In Fig.~\ref{deployment_PA}, we further evaluate the impact of the pinching-position deployment on the BER performance for the proposed DSM-PA scheme. Three classical deployment strategies are considered, where the pinching-position sets along each waveguide are given by $\{9,\,10,\,11\}~{\rm m}$, $\{2,\,10,\,18\}~{\rm m}$, and $\{1.5,\,7.5,\,18.5\}~{\rm{m}}$, corresponding to the closely spaced strategy, equally spaced strategy, and nonuniform spaced strategy, respectively. It can be observed that the nonuniform spaced deployment strategy  achieves the best BER performance, whereas the closely spaced deployment exhibits the worst performance, with the equally spaced configuration lying in between.

This behavior can be explained by the fact that the different pinching-position deployment strategies directly determine the distinguishability of the effective channel. When the pinching positions are concentrated in a narrow spatial interval, e.g., the closely spaced case, different activation patterns tend to induce highly correlated effective channels, thereby degrading the reliability of index detection. In contrast, the nonuniform spaced deployment strategy provides more distinguishable channel responses across different activated positions, which improves the index detection performance and reduces the overall BER. This result confirms the superiority of proper pinching-position deployment for improving the performance of the proposed DSM-PA scheme.

\vspace{-3mm}
\section{Conclusion}
In this paper, we proposed a differential spatial modulation scheme for PA systems to enable noncoherent transmission without instantaneous CSI.
By embedding both the Alamouti-based coding structure and the pinching-position index into the differential codeword, the proposed DSM-PA scheme effectively utilized the spatial degrees of freedom of the PA system whilst providing transmit diversity gain.
We derived an analytical upper bound on the BER using an MGF approach and showed that the proposed DSM-PA scheme achieves full diversity order.
Simulation results verified the accuracy of the derived BER analysis and demonstrated that the proposed scheme significantly outperforms the benchmark without Alamouti coding, while maintaining a reasonable BER loss with respect to the coherent scheme with perfect CSI.
The results also indicated that an appropriate deployment of pinching positions can further improve BER performance. 
In the future, we will focus on low-complexity detection for large-scale PA systems, optimizing deployment of pinching positions, and extending the DSM-PA framework to multi-user~scenarios.
\vspace{-2mm}

\bibliographystyle{IEEEtran}

\end{document}